\documentclass{article}
\usepackage{ijcai26}

\usepackage{times}
\usepackage{soul}
\usepackage{url}
\usepackage[hidelinks]{hyperref}
\usepackage[utf8]{inputenc}
\usepackage[small]{caption}
\usepackage{graphicx}
\usepackage{amsmath}
\usepackage{amsthm}
\usepackage{booktabs}
\usepackage{algorithm}
\usepackage{algorithmic}
\usepackage[switch]{lineno}

\usepackage{xcolor}
\usepackage{colortbl}
\usepackage{multirow}
\usepackage{amssymb}
\definecolor{gray}{RGB}{236,236,236}   
\definecolor{red}{RGB}{249,218,218}   
\definecolor{yellow}{RGB}{254,251,228} 
\definecolor{green}{RGB}{224,254,220}   
\definecolor{purple}{RGB}{217,217,252} 

\title{Representation-Regularized Convolutional Audio Transformer for Audio Understanding}

\author{
Bing Han$^{1\star}$\thanks{$^\star$Equal Contribution\quad$^\dagger$Corresponding Author}
\and
Chushu Zhou$^{1,2\star}$
\and
Yifan Yang$^{1}$
\and
Wei Wang$^{1}$
\and \\
Chenda Li$^{1}$
\and
Wangyou Zhang$^{1}$
\And
Yanmin Qian$^{1\dagger}$ \\
\affiliations
$^{1}$Auditory Cognition and Computational Acoustics Lab, Shanghai Jiao Tong University \\
$^{2}$Shanghai Innovation Institute\\
\emails
\{hanbing97, zhouchushu, yanminqian\}@sjtu.edu.cn
}

\begin{document}

\maketitle

\begin{abstract}
Bootstrap-based Self-Supervised Learning (SSL) has achieved remarkable progress in audio understanding. However, existing methods typically operate at a single level of granularity, limiting their ability to model the diverse temporal and spectral structures inherent in complex audio signals. Furthermore, bootstrapping representations from scratch is computationally expensive, often requiring extensive training to converge. In this work, we propose the Convolutional Audio Transformer (CAT), a unified framework designed to address these challenges. 
First, to capture hierarchical audio features, CAT incorporates a Multi-resolution Block that aggregates information across varying granularities. 
Second, to enhance training efficiency, we introduce a Representation Regularization objective. Drawing inspiration from generative modeling, this auxiliary task guides the student model by aligning its predictions with high-quality semantic representations from frozen, pre-trained external encoders. Experimental results demonstrate that CAT significantly outperforms baselines on audio understanding benchmarks. Notably, it achieves competitive performance on the AudioSet 20k dataset with 5 times faster convergence than existing methods. Codes and checkpoints will be released soon at \url{https://github.com/realzhouchushu/CAT}.
\end{abstract}

\section{Introduction}

Self-supervised learning (SSL) has become a cornerstone of representation learning, enabling models to extract meaningful patterns from unlabeled data across various modalities. In the audio domain, SSL has successfully mitigated the reliance on large-scale manual annotations, producing transferable representations for downstream tasks. Among these, bootstrap-based approaches, which employ a teacher-student framework to predict latent representations—have shown remarkable efficacy. State-of-the-art (SOTA) models such as data2vec series~\cite{data2vec,data2vec2}, EAT~\cite{eat}, ATST~\cite{atst} and M2D~\cite{m2d} leverage this paradigm to achieve impressive results in audio understanding tasks.

Despite these advancements, current bootstrap-based audio SSL methods face two critical limitations. First, they often overlook the inherent hierarchical nature of audio signals. Audio events span diverse temporal and spectral scales, ranging from transient acoustic textures to long-term semantic contexts. Existing methods typically process audio at a single, fixed level of granularity, limiting their ability to capture these multi-scale structures effectively. Second, training these models from scratch is inherently inefficient. Relying solely on internal consistency between the teacher and student networks requires extensive computational resources and prolonged training periods to bootstrap high-quality representations from random initialization.

\begin{figure*}[ht]
\centerline{\includegraphics[width=1.8\columnwidth]{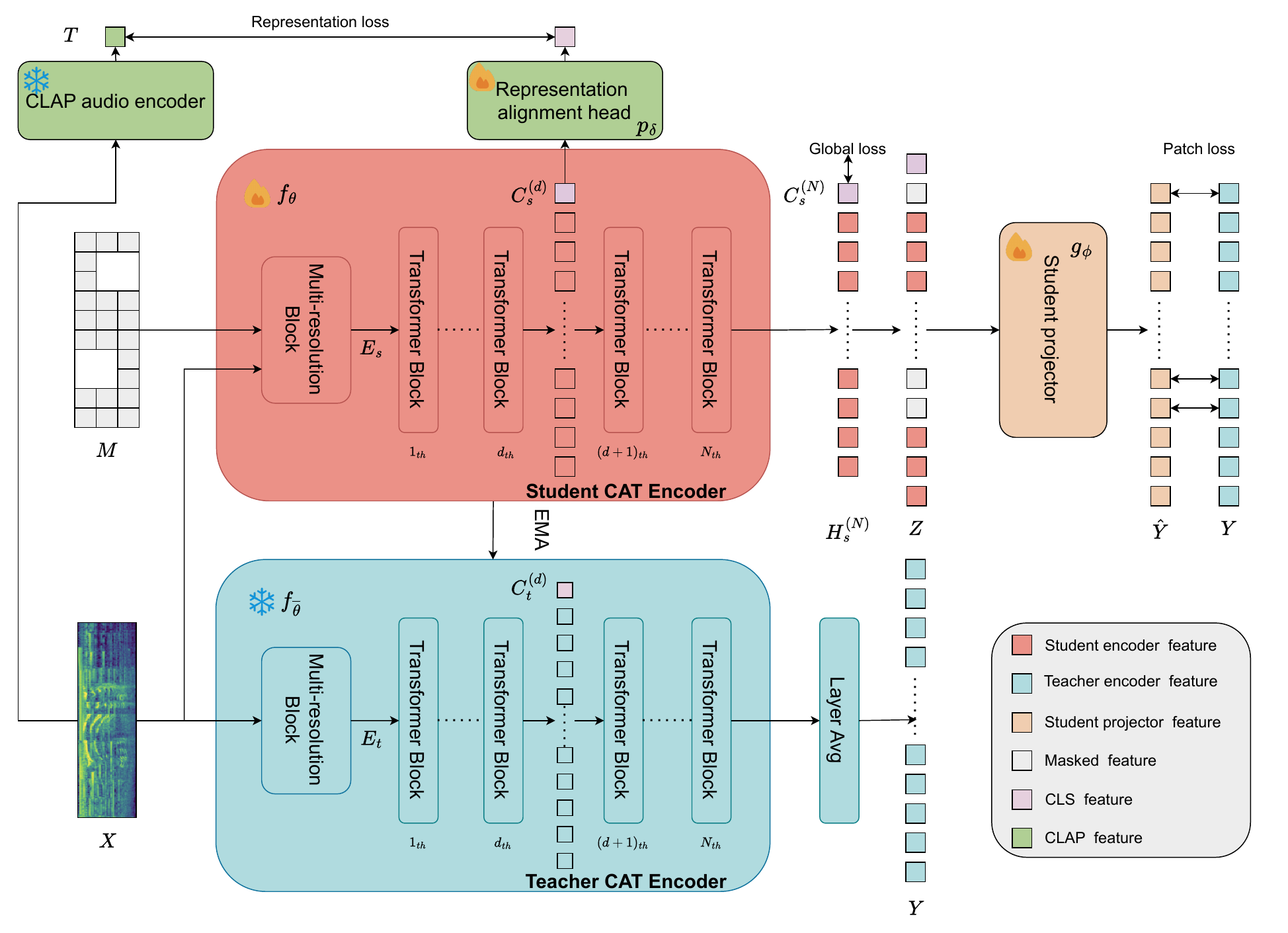}}
\caption{\textbf{CAT Pre-training Architecture Overview.} 
The model follows a student-teacher bootstrap paradigm. The Student Encoder processes a masked spectrogram, while the Teacher Encoder (updated via Exponential Moving Average, EMA) receives the unmasked input. The training objective is composed of three parts:
(1) Patch-level Loss ($L_p$): The student projector predicts the teacher's latent representations for masked patches;
(2) Global Loss ($L_g$): Aligns the global CLS token of the student with the teacher's aggregated features;
(3) Representation Loss ($L_r$): A regularization term that aligns intermediate representations from the student encoder with high-quality features extracted from a frozen external audio encoder.
}
\label{fig:CAT}
\end{figure*}

To address the efficiency bottleneck, we draw inspiration from recent advances in generative modeling, REPA~\cite{repa}. These studies demonstrate that generative models (e.g., diffusion models~\cite{ddpm,score}) converge significantly faster and generate higher-quality content when guided by representations from pre-trained encoders. We argue that the core objective of bootstrap SSL, predicting masked representations is fundamentally an implicit generative task. Just as diffusion models benefit from perceptual guidance, the ``reconstruction'' process in SSL can be accelerated by leveraging external semantic knowledge. Instead of learning solely from scratch, the student model can benefit from ``standing on the shoulders of giants'' by aligning with mature representations from external models.

Guided by this insight, we propose the Convolutional Audio Transformer (CAT), a unified framework designed to enhance both the granularity and efficiency of audio representation learning. To tackle the granularity issue, CAT incorporates a Multi-resolution Block. Unlike traditional single-layer patch embeddings, this module utilizes hierarchical convolutional layers to extract and aggregate features at multiple temporal and frequency scales, aligning with the multi-scale characteristics of audio signals. Simultaneously, to improve training efficiency, we introduce a Representation Regularization objective. We treat the student branch as a generative generator and align its intermediate features with high-quality representations from frozen, well-pretrained external encoders (e.g., CLAP~\cite{clap}, Audio-MAE~\cite{audio-mae}, AST~\cite{ast}). This auxiliary task provides stable semantic guidance, effectively "shortcutting" the early stages of representation learning.

Our experimental evaluations on audio and speech datasets validate the effectiveness of this approach. CAT not only achieves superior performance on audio understanding tasks compared to existing baselines  but also demonstrates remarkable efficiency, highlighting the power of combining multi-resolution processing with representation regularization.

Our main contributions are summarized as follows:
\begin{itemize}
\item We propose the Convolutional Audio Transformer (CAT), which replaces the standard patch embedding with a Multi-resolution Block to capture audio information across varying granularities.
\item We introduce Representation Regularization to audio SSL, formulating the masked prediction task as a generative process guided by external pre-trained encoders. This strategy significantly improves representation quality and training stability.
\item We demonstrate that CAT establishes new state-of-the-art (SOTA) results on standard audio understanding benchmarks with 5 times training speed-up.
\end{itemize}

\section{Related Work}
\subsection{Pre-trained Audio Encoder}
Supervised pre-training, contrastive pre-training and self-supervised pre-training are the three widely adopted approaches for audio encoder pre-training.
Supervised pre-training usually relies on collected labels to guide model training.
Representations extracted from supervised learning methods such as PANN~\cite{pann}, AST~\cite{ast} and HTS-AT~\cite{hts-at} perform well on various tasks. 
For contrastive pre-training, CLAP~\cite{clap} consists of dual encoders—one for audio and one for text—trained jointly to align audio representations with their corresponding natural language descriptions.
However, these methods typically require large amounts of labeled data, which limits their scalability.
Self-supervised learning (SSL) methods have recently gained more attention, as they can learn high-quality representations while eliminating the need for labeled data.
Some self-supervised methods including MAE-AST~\cite{mae-ast}, Audio-MAE~\cite{audio-mae} and MaskSpec~\cite{maskspec} follow the fashion of MAE~\cite{mae}, using features of unmasked patches to predict masked audio patches.
Another popular type of methods is bootstrap-based self-supervised methods. 
These approaches such as data2vec~\cite{data2vec}, EAT~\cite{eat} and A-JEPA~\cite{ajepa} adopt a student-teacher framework where the student model learns to predict representations generated by a slowly updated teacher model.
Bootstrap-based self-supervised methods have demonstrated promising performance on various audio understanding tasks.

\subsection{Representation Learning as Auxiliary Tasks}

Beyond traditional training paradigms, recent studies in generative modeling have explored using external representations to guide the learning process. 
In the domain of image generation, REPA~\cite{repa} demonstrated that diffusion models—which generate images from noise—converge significantly faster when their intermediate features are regularized to align with representations from pre-trained encoders (e.g., DINOv2~\cite{dinov2}). Subsequent works have extended this concept: iREPA~\cite{irepa} emphasizes the importance of spatial structural information, SARA~\cite{sara} introduces multi-level representation alignment to optimize the diffusion training process and accelerate convergence, SoftREPA~\cite{softrepa} applies it to multimodal text-to-image generation, and Semantic-VAE~\cite{semanticvae} utilizes alignment to enhance speech synthesis quality. 
We draw a parallel between these generative tasks and bootstrap-based SSL. We argue that the core objective of Masked SSL—predicting masked representations from partial inputs—is effectively an implicit generative task. Despite this theoretical connection, its potential to accelerate convergence and improve feature quality in understanding tasks (e.g., audio classification) remains underexplored. Our work bridges this gap by integrating representation regularization into the SSL framework, leveraging diverse external pretrained encoders to provide additional guidance for efficient audio understanding.

\section{Method}

\subsection{Model Architecture Overview}
CAT is a pre-trained model following the bootstrap paradigm, similar to established methods such as data2vec~\cite{data2vec}, DINO~\cite{dinov2} and EAT~\cite{eat}.
The model architecture of CAT is illustrated in Fig.~\ref{fig:CAT}. 
CAT employs a dual-branch architecture encompassing a student model and a teacher model.
The student model contains a CAT encoder $f_{\theta}$ followed by a projector module $g_{\phi}$, while the teacher model solely consists of a CAT encoder $f_{\bar{\theta}}$. 
Although the student and teacher encoders share an identical architecture, they do not share parameters. 
This asymmetric design, with the projector only on the student side, effectively prevents representation collapse (i.e., the encoder outputting constant representations regardless of the input data).  

During pre-training, CAT takes an audio spectrogram $\mathbf{X} \in \mathbb{R}^{T \times F}$ as input, where $T$ and $F$ denote the time and frequency dimensions, respectively. 
The model employs two sets of hyperparameters: resolution parameters ${r_1, r_2, \dots, r_{n_r}}$ and feature channel dimensions ${D_1, D_2, \dots, D_{n_r}}$, where $n_r$ is the number of layers in the multi-resolution block, with each $r_i$ $(i = 1, \dots, n_r)$ specifying a distinct non-overlapping patch size for the $i$-th resolution layer, while $D_i$ denotes the number of feature channels in the corresponding layer.
Let $P$ denote the total number of patches, where $P = {TF}/{r_{n_r}^2}$.
In the teacher branch, the encoder $f_{\bar\theta}$ processes the entire spectrogram through a multi-resolution block, which subsequently outputs a feature representation $\mathbf{E_t} \in \mathbb{R}^{P \times D_{n_r}}$ that integrates information from multiple granularities.
A comprehensive description of how the multi-resolution block works is provided in \ref{subsec:cat}.
Then, a learnable CLS token is added to this feature sequence for global representation learning.
This representation is then fed into a stack of transformer blocks to further capture contextual dependencies.
We denote the output of the $j$-th transformer block as $\mathbf{H_t^{(j)}}$ for the teacher branch and $\mathbf{H_s^{(j)}}$ for the student branch $(j = 1, \dots, N)$, where $N$ is the number of transformer blocks. 

The student and teacher models share an identical encoder architecture but differ slightly in their input processing.
For the teacher encoder, the target $Y$ is derived from a mean-pooling operation over the hidden states of its internal layers.
\begin{equation}
\mathbf{Y}=\sum^{N}_{i=1}\frac{\mathbf{H_t^{(i)}}}{N}
\end{equation}

For the student encoder, the student multi-resolution block takes an additional mask as input $\mathbf{M} \in \mathbb{R}^{P \times D_{n_r}}$. 
This mask is applied to the patchified spectrogram, yielding features $\mathbf{E_s}$.
To enhance training efficiency, only the features corresponding to these unmasked patches are processed and output by the transformer blocks.
After obtaining the output $\mathbf{H_s^{(N)}}$ from the final transformer block, pad embeddings are interpolated at the positions of the masked patches to form the projector input $\mathbf{Z}$.
The projector $g_{\phi}$ comprises a sequence of lightweight convolutional layers.
The feature $\mathbf{Z}$ is then fed into the student projector $g_{\phi}$, which outputs $\mathbf{\hat{Y}}$ that serves as the prediction of the output of the teacher encoder $\mathbf{Y}$.

\subsection{Convolutional Audio Transformer Encoder}
\label{subsec:cat}

\begin{figure}[htb]
\centerline{\includegraphics[width=\columnwidth]{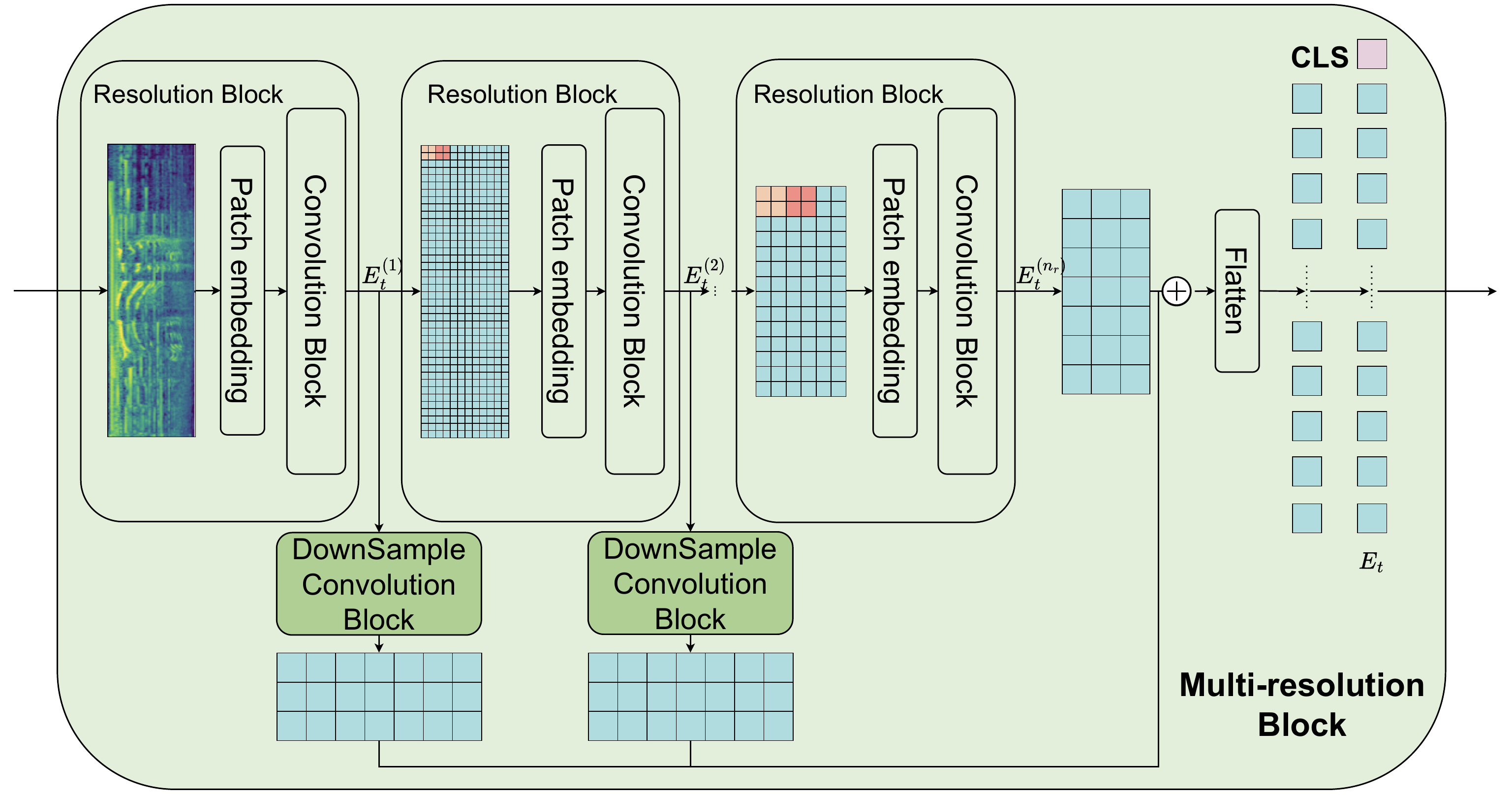}}
\caption{\textbf{Multi-Resolution Block Architecture.} This diagram depicts the data flow in the teacher's Multi-Resolution Block. The student branch follows a nearly identical process, differing primarily by the introduction of an input mask that is element-wise multiplied with the features inside the Convolutional Block.}
\label{fig:mul}
\end{figure}

The CAT encoder comprises a multi-resolution block for extracting features at different granularities, followed by a stack of $N$ transformer blocks. 
As shown in Fig.~\ref{fig:mul}, The multi-resolution block contains $n_r$ resolution blocks.

For the teacher CAT encoder, The first block takes the spectrogram $\mathbf{X} \in \mathbb{R}^{T \times F}$ as input. 
The output of the $k$-th resolution block, denoted as $\mathbf{E^{(k)}_t} \in \mathbb{R}^{\frac{T}{r_{k}} \times \frac{F}{r_{k}} \times D_{k}}$, serves as the input to the subsequent $(k+1)$-th block for $k = 1, \dots, n_r-1$. 
Given $\mathbf{E^{(k)}_t}$, the $(k+1)$-th resolution block extracts coarser-grained features using a patch embedding module, which is implemented as a convolutional layer with a stride and kernel size of $r_{k+1}/r_k$. 
This feature is then passed through a subsequent convolution module composed of several lightweight convolutional layers, which are designed to further integrate information while preserving the shape of the feature, yielding the output $\mathbf{E^{(k+1)}_t}$.
Since the output shapes of the different resolution blocks vary, several additional downsample blocks are introduced to project each intermediate representation $\mathbf{E^{(k)}_t}$ to the same shape as the final output $\mathbf{E^{(n_r)}_t} \in \mathbb{R}^{\frac{T}{r_{n_r}} \times \frac{F}{r_{n_r}} \times D_{n_r}}$. 
The downsampled outputs are summed together and then flattened.
Subsequently, a learnable classification (CLS) token is added to the beginning of this sequence to form the final integrated representation $\mathbf{E_t}$.

The student CAT encoder is almost identical to the teacher encoder, with the key difference that it accepts an additional mask $\mathbf{M}$ as input. 
This mask is interpolated and reshaped to match the spatial dimensions of each intermediate feature $\mathbf{E^{(k)}_s}$, and is then applied to the features within the convolutional block.
Another difference is that, for the student CAT encoder, all features corresponding to masked patches are removed, retaining only the unmasked features before these features are finally passed into transfomer blocks to reduce computation. 

\subsection{Training Objective with Representation Regularization}

Representation regularization has demonstrated promising performance on generative tasks. 
Benefiting from the asymmetric structure of CAT, we argue that the student branch (the student CAT encoder together with the student projector) is an implicit generative model. 
The key difference is that the target domain of its generation is not raw images or raw audio signals, but the teacher model’s representations corresponding to the masked patches.
Introducing representation regularization as an auxiliary task may improve the student model's generation performance (e.g., leading to better alignment with the teacher), and therefore enhance the representation learning quality of the encoder. 
Furthermore, guided by well-trained external representations, the training process can be easier compared with relying solely on the model to learn representations independently from scratch.

Motivated by this, we introduce representation regularization as an auxiliary task optimized jointly with the original main objective. The overall objective $L$ can be defined as:
\begin{equation}
L=L_p+\lambda_{1} L_g+\lambda_{2} L_r
\end{equation}
where $L_p$ and $L_g$ are the patch-level loss and global loss in the original main objective, and $L_r$ is the representation learning loss used for representation regularization.
$\lambda_1$ and $\lambda_2$ are hyperparameters manually assigned as the loss weights of $L_g$ and $L_r$.

We align the intermediate representations of our model with those extracted by an external model by introducing a representation learning loss $L_r$.
Specifically, we feed the unmasked audio into a frozen pre-trained encoder such as  CLAP~\cite{clap} to extract representations $T \in \mathbb{R}^{1\times D_t}$, which are then used as the target.
A single-layer representation alignment head $p_{\delta}$ operates on the CLS token of the $d$-th student transformer block ($d$ is a hyperparameter), mapping its feature dimension from $D_{n_r}$ to $D_t$.
The $L_r$ can be given by:
\begin{equation}
L_r = \left\| \mathbf{p_{\delta}(C_s^{(d)})} - \mathbf{T} \right\|_2^2
\end{equation}

This strategy encourages the model to learn global semantic information that is invariant to patch masking while providing meaningful guidance to accelerate convergence.

The original training objectives are retained following the setting of Data2vec~\cite{data2vec2}.
To compute the patch-level loss $L_p$, the student projector takes the unmasked patch features $H_s^{(N)}$ from the student encoder and outputs $\mathbf{\hat{Y}}$ to predict the teacher encoder's outputs $\mathbf{Y}$ for both masked and unmasked patches. 
$L_p$ is computed only over the masked patches, formulated as the Mean Squared Error (MSE) between the prediction and the target:
\begin{equation}
L_p = \left\| \mathbf{\hat{Y}_{masked}} - \mathbf{Y_{masked}} \right\|_2^2
\end{equation}

To capture the global audio representation, we use the CLS token from the final student transformer output $\mathbf{C_s^{(N)}} \in \mathbb{R}^{1 \times D_{n_r}}$ to predict the mean-pooled hidden states of the teacher transformer blocks, denoted as $\mathbf{\bar{C}_t} \in \mathbb{R}^{1 \times D_{n_r}}$. The global loss $L_g$ is formulated as:
\begin{equation}
L_g = \left\| \mathbf{C_s^{(N)}} - \mathbf{\bar{C}_t} \right\|_2^2
\end{equation}

During the training process, the student CAT encoder $f_{\theta}$, the projector $g_{\phi}$ and the representation alignment head $p_{\delta}$ are trainable, while the teacher CAT encoder $f_{\bar{\theta}}$ is frozen. 
The teacher CAT encoder is updated using an Exponential Moving Average (EMA) strategy with a linearly increasing momentum coefficient $\tau$. 
This process is described by the following equation:
\begin{equation}
\bar{\theta}\leftarrow \tau \theta + (1-\tau)\bar{\theta}
\end{equation}

\section{Experiments}

\begin{table*}[htb]
\centering
\caption{\textbf{Comparison of supervised and self-supervised pre-trained models.} Pre-training data abbreviations: ImageNet (IN), AudioSet (AS) and LibriSpeech (LS). TA and TI denote the 128K text-audio pairs and 400M text-image pairs for CLAP and CLIP pre-training, respectively.}

\begin{tabular}{lcccccc}
\toprule
\multirow{2}{*}{\textbf{Model}} & \multirow{2}{*}{\textbf{\#Param}} & \textbf{Pre-training} & \textbf{AS-2M} & \textbf{AS-20K} & \textbf{ESC-50} & \textbf{SPCV2} \\
 & & \textbf{Data} & \textbf{mAP(\%)} & \textbf{mAP(\%)} & \textbf{Acc(\%)} & \textbf{Acc(\%)}\\

\midrule
PANN~\cite{pann} & 81M & - & 43.1 & 27.8 & 83.3 & 61.8\\
PSLA~\cite{psla} & 14M & IN & 44.4 & 31.9 & - & 96.3\\
AST~\cite{ast} & 86M & IN & 45.9 & 34.7 & 88.7 & 98.1\\
MBT~\cite{mbt} & 86M & IN-21K & 44.3 & 31.3 & - & -\\
PassT~\cite{paast} & 86M & IN & 47.1 & - & 96.8 & -\\
HTS-AT~\cite{hts-at} & 31M & IN & 47.1 & - & 97.0 & 98.0\\
Wav2CLIP~\cite{wav2clip} & 74M & TI+AS & - & - & 86.0 & -\\
AudioCLIP~\cite{audioclip} & 93M & TI+AS & 25.9 & - & 96.7 & -\\
Conformer~\cite{conformer} & 88M & AS & 41.1 & - & 88.0 & - \\
SS-AST~\cite{ssast} & 89M & AS+LS & - & 31.0 & 88.8 & 98.0\\
MAE-AST~\cite{mae-ast} & 86M & AS+LS & - & 30.6 & 90.0 & 97.9\\
MSM-MAE~\cite{msm-mae} & 86M & AS & - & - & 85.6 & 87.3\\
data2vec~\cite{data2vec} & 94M & AS & - & 34.5 & - & -\\
Audio-MAE~\cite{audio-mae} & 86M & AS & 47.3 & 37.1 & 94.1 & 98.3\\
BEATs$_{\text{iter3}}$~\cite{beats} & 90M & AS & 48.0 & 38.3 & 95.6 & 98.3\\
BEATs$_{\text{iter3+}}$~\cite{beats} & 90M & AS & 48.6 & 38.9 & 98.1 & 98.1\\
CLAP~\cite{clap} & 86M & TA & 46.9 & 36.7 & 96.7 & 96.8 \\
MaskSpec~\cite{maskspec} & 86M & AS & 47.1 & 32.3 & 89.6 & 97.7\\
A-JEPA~\cite{ajepa} & 86M & AS & 48.6 & 38.4 & 96.3 & 98.5\\ 
EAT~\cite{eat} & 88M & AS & 48.6 & 40.2 & 95.9 & 98.3\\
M2D-CLAP~\cite{m2d-clap} & 86M & TA+AS &  48.5 & 41.8 & 97.4 & 98.3\\
ATST-Frame~\cite{atst} & 86M & AS & 48.0 & 39.0 & - & 98.1\\
ATST-Clip~\cite{atst} & 86M & AS & 45.2 & 37.9 & - & 98.0\\
ASiT~\cite{asit} & 86M & AS & 48.0 & 38.6 & 95.3 & \textbf{98.9} \\
ASDA~\cite{asda} & 93M & AS & 49.0 & 41.5 & 96.1 & 98.3 \\
\midrule
\multicolumn{5}{l}{\textit{\textbf{Ours}}} \\
CAT & 91M & AS & \textbf{50.2} & \textbf{47.8} & \textbf{98.6} & 98.3 \\
\bottomrule
\end{tabular}
\vspace{-0.2cm}
\label{tab:pretrain_comparison}
\end{table*}

\subsection{Datasets}

We conduct self-supervised pre-training on the AS-2M subset of AudioSet~\cite{audioset}, utilizing approximately 1.9 million audio clips without labels. For downstream evaluation, we benchmark our model on three datasets: AudioSet (fine-tuning on both the full AS-2M and the balanced AS-20K subsets), ESC-50~\cite{esc-50}, and Speech Commands V2 (SPC-2)~\cite{spc-2}. We report mean Average Precision (mAP) for AudioSet tasks. For ESC-50, we assess out-of-domain generalization using standard 5-fold cross-validation accuracy. Finally, we evaluate speech understanding capabilities on SPC-2, employing the official training/test splits and reporting classification accuracy.

More details are provided in Appendix.

\subsection{Experimental Setup}
\textbf{Input Processing} To extract the input audio spectrogram, each audio clip is first resampled to 16 kHz. 
We then compute the Mel-frequency spectrogram using 128 filter banks, a 25-ms Hanning window, and a 10-ms frame shift. 
For the masking strategy, we adopt the Inverse Block Masking proposed in Data2Vec 2.0~\cite{data2vec2}, with a mask ratio of 80\%. This strategy begins by masking all patches and then iteratively unmasks blocks of adjacent patches until the desired masking ratio is reached.
For data augmentation, we create four distinct masks for each audio clip.

\noindent \textbf{Model Architecture} The CAT encoder begins with the Multi-resolution Block, which progressively projects input patches into a 768-dimensional feature space via convolutional layers. This is followed by a Transformer backbone consisting of 12 layers with a hidden dimension of 768 and 12 attention heads. To ensure fair comparisons, we adjust the number of transformer layers in specific ablation settings to maintain consistent parameter counts with baselines. The student projector consists of five strided convolutional layers followed by a linear projection.

\noindent \textbf{Training Details} Pre-training is conducted on AS-2M for 400k steps with a batch size of 96 using 2 NVIDIA H200 GPUs, taking approximately 3 days. We use the Adam optimizer ($\beta_1=0.9, \beta_2=0.95$) with a cosine annealing scheduler~\cite{sgdr}, a peak learning rate of $5\times10^{-4}$, and a warm-up phase of 53,333 steps. The global loss weight $\lambda_1$ is set to 1.0. For supervised fine-tuning, models are trained for 300k steps on AS-2M and 40k steps on other datasets.

More details are provided in Appendix.

\subsection{Main Results}

We compare CAT with state-of-the-art (SOTA) baselines across three distinct training paradigms: supervised learning (e.g., AST, HTS-AT), contrastive learning (e.g., CLAP), and self-supervised learning (e.g., Audio-MAE, EAT, BEATs). The comprehensive results are reported in Table~\ref{tab:pretrain_comparison}.
On AudioSet, CAT establishes a new SOTA result, achieving 50.2 mAP on AS-2M and 47.8 mAP on AS-20K. Notably, it outperforms the leading bootstrap method ASDA by +1.2 mAP on AS-2M and by a remarkable +6.3 mAP on the smaller AS-20K dataset. These results validate that our Multi-resolution Block effectively captures complex hierarchies, while Representation Regularization acts as a semantic scaffold to significantly boost data efficiency. CAT also demonstrates superior generalization, setting a new benchmark on ESC-50 with 98.6\% accuracy and maintaining competitive performance on Speech Commands V2 (98.3\%), confirming its robustness across diverse audio and speech domains.

\subsection{Pre-training Efficiency}
\begin{figure}[htbp]
    \centering
    \includegraphics[width=\linewidth]{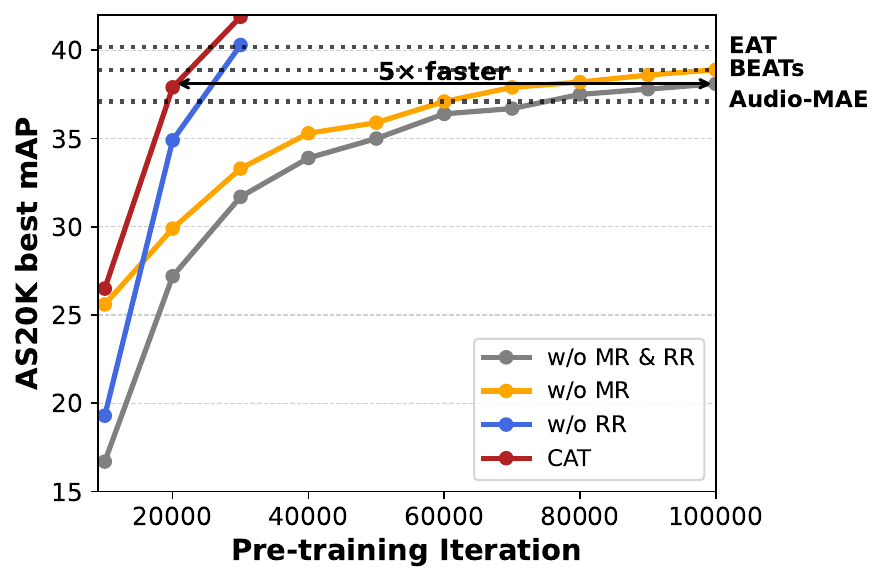}
    \caption{\textbf{CAT Pre-training Convergence Performance.} ``w/o MR'' denotes the model using a vanilla patch embedding instead of the multi-resolution block, while ``w/o RR'' denotes the model without representation regularization. The x-axis indicates the pre-training step (showing only the early phase), while the y-axis shows the best mAP on AS-20K after supervised fine-tuning with the corresponding checkpoint. For ease of comparison, the best reported performances of EAT, BEATs and Audio-MAE models are indicated by horizontal dashed lines in the plot.}
    \label{fig:pre-training-convergence}
\end{figure}

We analyze the convergence behavior of CAT in Figure~\ref{fig:pre-training-convergence}. The results demonstrate that both the architectural design and the regularization objective contribute significantly to training efficiency.
First, the Multi-resolution Block introduces effective structural inductive biases. Compared to the single-resolution baseline (``w/o MR''), this hierarchical design enables the model to capture discriminative audio patterns much more rapidly in the early training stages. Second, Representation Regularization acts as a powerful accelerator. As shown by the comparison with the ``w/o RR'' variant, incorporating external semantic guidance induces a significantly steeper learning curve. This confirms that guiding the student with a mature teacher effectively mitigates the slow ``cold-start'' phase inherent in bootstrapping from scratch. By combining these two innovations, CAT achieves a competitive 37.9\% mAP on AS-20K within just 20,000 iterations. In contrast, baseline methods typically require 100,000 iterations to reach comparable performance. This translates to a 5$\times$ improvement in pre-training efficiency, significantly reducing the computational cost of developing high-performance audio models.



\subsection{Ablation Study}
\begin{table}[htb]
\centering
\caption{\textbf{Component-wise ablation.}  ``w/o MR'' denotes the model using a vanilla patch embedding instead of the multi-resolution block, while ``w/o RR'' denotes the model without representation regularization. }
\begin{tabular}{lcccc}
\toprule
\textbf{Model} & \textbf{AS2M} & \textbf{AS20K} & \textbf{ECS50} & \textbf{SPCV2} \\
\midrule
\rowcolor{gray}CAT & \textbf{50.2} & \textbf{47.8} & \textbf{98.6} &\textbf{98.3} \\
w/o MR & 50.0 & 47.3 & 98.6 & 98.1\\
w/o RR  & 49.4 & 41.6 & 96.5 & 98.3 \\
w/o MR\&RR & 48.9 & 40.6 & 96.5 & 98.0 \\
\bottomrule
\end{tabular}
\label{tab:component}
\end{table}
\subsubsection{Ablation Study of Component}


We perform a component-wise ablation in Table~\ref{tab:component} to isolate the contributions of the Multi-resolution (MR) Block and Representation Regularization (RR). The results highlight the critical role of Representation Regularization in boosting data efficiency. Removing this component (``w/o RR'') results in a drastic performance drop, particularly on the data-limited AS-20K benchmark, where mAP plummets from 47.8\% to 41.6\%. This substantial 6.2\% gap confirms our hypothesis: external semantic guidance acts as a crucial scaffold, enabling the model to learn high-quality representations even when pre-training data is limited. Similarly, the Multi-resolution Block proves indispensable for capturing fine-grained audio patterns. Replacing our hierarchical block with a standard patch embedding (``w/o MR'') consistently degrades performance across both AS-2M and AS-20K datasets. Ultimately, the full CAT model achieves the best performance, demonstrating that hierarchical feature extraction and generative semantic regularization are complementary mechanisms that jointly drive the model's state-of-the-art results.

\subsubsection{Ablation Study of Multi-Resolution Block}
\begin{table}[htb]
\centering
\caption{\textbf{Ablation of Resolution Configurations.} The multi resolution column indicates the set of resolution hyperparameters $\{r_1, r_2, \dots r_{n_r}\}$ used in the multi-resolution block. The transformer \#layer column specifies the number of Transformer blocks. }
\begin{tabular}{lccc}
\toprule
\multicolumn{1}{c}{\textbf{Multi}} & \textbf{Transformer} & \multirow{2}{*}{\textbf{\#Param}} & \textbf{AS-20K} \\
\multicolumn{1}{c}{\textbf{Resolution}} & \textbf{\#Layer} &  & \textbf{mAP(\%)} \\
\midrule
 \{16\} & 12 & 88M & 40.2 \\ 
\{4,16\} & 12 & 100M & 41.4 \\ 
\{8,16\} & 12 & 93M &  \textbf{41.6} \\ 
\{4,8,16\} & 11 & 91M & \textbf{41.6} \\ 
\{4,8,16,32\} & 12 & 128M & 38.1 \\ 
\bottomrule
\end{tabular}
\label{tab:conv_ab}
\end{table}

The number of resolution blocks and the patch resolution adopted for these blocks are the most important parameters for the multi-resolution block configuration. 
Table~\ref{tab:conv_ab} investigates the effects of different resolution block configurations within the multi-resolution block.
Since all features extracted from the different resolution blocks are downsampled to the same shape as the final resolution block via the subsequent downSample convolution block, the final resolution determines the sequence length of the features input to the subsequent transformer blocks.
For fair comparison, the final resolution in most settings is set to 16, following existing works. Extremely large resolution sets (e.g., $\{{4,8,16,32}\}$) significantly reduce the feature length, leading to suboptimal performance.
It is worth mentioning that the single-resolution setting $\{{16}\}$ essentially degenerates to the baseline model, but with an additional convolution block appended.
As demonstrated in Table~\ref{tab:conv_ab}, the results indicate that both the set of resolution hyperparameters ${r_i}$ and the model depth significantly impact performance.
The setting $\{{4,8,16}\}$, using 3 resolution blocks with resolutions 4, 8, and 16 separately, achieves the best performance.
We adopt $\{{4,8,16}\}$ as our default setting, as it achieves  performance comparable to the $\{8,16\}$ setting with fewer model parameters. 
It is worth noting that for the $\{{4,8,16}\}$ setting, one transformer layer is removed to ensure the model size remains comparable to that of existing methods.


\subsubsection{Ablation Study of Representation Regularization}

\begin{table}[htb]
\centering
\caption{\textbf{Ablation of Target representation.} The Model column indicates the model from which representations are extracted. We utilize the hidden states from the last layer for representation regularization. The first row is the baseline without using representation regularization  for comparison.}
\resizebox{\linewidth}{!}{
\begin{tabular}{ccccccc}
\toprule
\textbf{Model} & \textbf{AS2M} & \textbf{AS20K} & \textbf{ECS50} & \textbf{SPCV2} & \textbf{Avg.}\\
\midrule
\rowcolor{gray}-  & 48.6 & 40.2 & 95.9 & 98.3 & 70.8 \\ \midrule
\cellcolor{red}CLAP & 50.0 & \textbf{47.3} & 98.6 & \textbf{98.3} & \textbf{73.6} \\
\cellcolor{red}AST & 50.0 & 46.6 & \textbf{98.7} & 98.1 & 73.4\\
\cellcolor{red}AudioMAE & \textbf{50.6} & 46.6 & 98.4 & 97.9 & 73.4 \\
\bottomrule
\end{tabular}
}
\label{tab:representation}
\end{table}
\colorbox{red}{\textbf{Target representation Source.}} To investigate how different types of representations affect the training process and identify the optimal target representation, we conduct ablation studies using representations extracted from models trained in different manners.
We find that representation regularization generalizes well, as shown in Table~\ref{tab:representation}. 
For models pre-trained with representation regularization, we observe substantial improvements over the baseline on general audio related metrics, but a slight drop on SPCV2 (except for the model guided by CLAP features).
We suspect that other representations are pre-trained only on audio data, which renders them strong at audio understanding but limits their generalization to speech understanding.
CLAP is used as the target representation unless otherwise mentioned in the following experiments, as it yields the best overall performance.


\begin{table}[htb]
\caption{\textbf{Component-wise analysis.} The aligned layer $d$ specifies which student transformer block is aligned, the regularization loss weight is $\lambda_2$ for regularization loss term $L_r$.}
\label{sample-table}
\begin{center}
\begin{tabular}{cccc} 
\toprule
 \textbf{Aligned} & \textbf{Regularization} & \multirow{2}{*}{\textbf{Objective}} & \textbf{AS-20K} \\
 \textbf{Layer} & \textbf{Loss Weight} &  & \textbf{mAP(\%)} \\
\midrule
\cellcolor{yellow}12 & 1.0 & $\mathrm{MSE}$ & \textbf{47.3} \\
\cellcolor{yellow}10 & 1.0 & $\mathrm{MSE}$ & 47.1 \\
\cellcolor{yellow}8 & 1.0 & $\mathrm{MSE}$ & 46.6 \\
\midrule
12 & \cellcolor{green}5.0 & $\mathrm{MSE}$ & 47.1  \\
12 & \cellcolor{green}1.0 & $\mathrm{MSE}$ & \textbf{47.3}  \\
12 & \cellcolor{green}0.1 & $\mathrm{MSE}$ & 45.8  \\
\midrule
12 &1.0 &\cellcolor{purple}$\mathrm{MSE}$ & \textbf{47.3} \\
12 &1.0 &\cellcolor{purple}$\mathrm{CE}$ & 42.4 \\
12 &1.0 &\cellcolor{purple}$\mathrm{l}_1$ & 46.8 \\
12 &1.0 &\cellcolor{purple}$\mathrm{cosine}$ & 47.2 \\
\bottomrule
\end{tabular}
\vspace{-0.2cm}
\end{center}
\label{tab:rr_ab}
\end{table}

\colorbox{yellow}{\textbf{Representation Regularization Layer.}}
The choice of which student transformer block to align with the CLAP representation significantly impacts performance. 
Applying the representation regularization loss $L_r$ to the CLS token from the final student block yields the best result. 
Performance gradually degrades when the alignment is applied to earlier student blocks. 
This trend indicates that the feature representations within the student model become progressively more refined and semantically meaningful through the layers. 
Aligning the high-level representations from student blocks with the powerful CLAP embedding $T$ provides a more effective and stable training signal for the overall model.

\colorbox{green}{\textbf{Loss Weight.}} The weighting coefficient $\lambda_2$ for $L_r$ balances its influence against other loss terms. A weight of 1.0 yields the optimal result. 
Deviating from this value—either increasing to 5.0 or decreasing to 0.1 —leads to a performance drop. 
This confirms that representation regularization plays a crucial role.
An overly strong weight may distort the primary pre-training task, while an overly weak one provides insufficient guidance.

\colorbox{purple}{\textbf{Objective Function.}} 
The choice of the objective function is critical for effective feature alignment. 
The Mean Squared Error (MSE) loss achieves the best performance, significantly outperforming the Cross-Entropy (CE) loss and marginally surpassing both the $l_1$ loss and the Cosine similarity loss.

\subsubsection{Audio Caption Understanding}
To further assess the semantic richness of the learned representations, we evaluate CAT and baseline systems on the Audio Captioning task and results are provided in Appendix.

\section{Conclusion}

In this work, we proposed the Convolutional Audio Transformer (CAT) to address the limitations of single-granularity processing and inefficient training in bootstrap-based SSL. We introduced a Multi-resolution Block to capture hierarchical audio structures and leveraged Representation Regularization, which is inspired by generative modeling, to guide the learning process with external semantic priors. By treating masked prediction as an implicit generative task, our approach significantly accelerates convergence while enhancing feature quality. Experiments confirm that CAT establishes new state-of-the-art performance on AudioSet and ESC-50 while achieving 5× faster convergence than existing baselines, validating the potential of bridging self-supervised learning with generative guidance for efficient audio understanding.

\newpage
\appendix

\section*{Ethical Statement}
There are no ethical issues. All data used in this work were collected and processed in accordance with relevant ethical guidelines and licensing terms.

\bibliographystyle{named}
\bibliography{ijcai26}

\begin{table*}[htb]
  \centering
  \caption{CAT pre-training and fine-tuning hyper-parameters.}
  \label{tab:hyperparams}
    \begin{tabular}{lcccccc}
      \toprule
      \multirow{2}{*}{Hyperparameters} & Pre-Training & \multicolumn{4}{c}{Fine-Tuning} \\
      \cmidrule(r){3-6}
      & AS-2M & AS-2M & AS-20K & ESC-50 & SPCV2 \\
      \midrule
      Optimizer & \multicolumn{5}{c}{AdamW~\cite{adamw}} \\
      Optimizer Momentum & \multicolumn{5}{c}{$\beta_1=0.9, \beta_2=0.95$} \\
      Weight Decay & \multicolumn{5}{c}{0.05} \\
      Learning Rate Schedule & \multicolumn{5}{c}{Cosine~\cite{sgdr}} \\
      Peak Learning Rate & 0.0005 & 0.00005 & 0.00005 & 0.00005 & 0.0002 \\
      Minimum Learning Rate & \multicolumn{5}{c}{0.000001} \\
      Steps & 400K & 300K & 40K & 4K & 40K \\
      Warm-up steps & 53K & 30K & 4K & 400 & 4K \\
      Batch size & 12 & 96 & 48 & 48 & 256 \\
      Clone batch & 16 & \multicolumn{4}{c}{N/A} \\
      Number of GPUs & 4 & \multicolumn{4}{c}{1} \\
      Dropout~\cite{dropout} & 0.0 & 0.0 & 0.0 & 0.0 & 0.0 \\
      Drop path~\cite{droppath} & 0.0 & 0.1 & 0.1 & 0.1 & 0.1 \\
      Weighted Sampling & False & True & False & False & False \\
      Weighted Sampling size & N/A & 200K & N/A & N/A & N/A \\
      Roll Augmentation & False & True & True & True & False \\
      Noise Augmentation & False & False & False & False & True \\
      SpecAug~\cite{specaugment} & N/A & 0.2 & 0.2 & 0.2 & 0.1 \\
      Mixup~\cite{mixup} & 0.0 & 0.8 & 0.8 & 0.0 & 0.8 \\
      Multi-label & N/A & True & True & False & False \\
      Loss Function & MSE & BCE & BCE & CE & BCE  \\
      Dataset Mean for Normalization & -4.268 & -4.268 & -4.268 & -6.627 & -6.846 \\
      Dataset Std for Normalization & 4.569 & 4.569 & 4.569 & 5.359 & 5.565 \\
      \bottomrule
    \end{tabular}
\end{table*}

\section{Datasets Detail}
We utilized AudioSet for self-supervised pre-training and evaluated the learned representations on AudioSet, ESC-50, and Speech Commands V2 through supervised fine-tuning. The details of these datasets are as follows:

\noindent \textbf{AudioSet}
AudioSet~\cite{audioset} is a large-scale audio event dataset consisting of approximately 2 million 10-second audio clips extracted from YouTube videos, covering 527 distinct audio event classes. The dataset is organized into three subsets: the unbalanced training subset (AS-2M), the balanced training subset (AS-20K), and the evaluation subset. We use the huggingface version\footnote{https://huggingface.co/datasets/confit/audioset-qiuqiangkong}, which contains $20,550/22,160$ of the balaned training subset, $1,913,637 / 2,041,789$ of the unbalanced training subset, and $18,887 / 20,371$ of the evaluation subset. In this work, we use the AS-2M subset for pre-training, utilizing only the audio data without labels. For downstream evaluation, we conduct supervised fine-tuning on both AS-2M and AS-20K and report the mean Average Precision (mAP) on the official evaluation subset.

\noindent \textbf{Environmental Sound Classification (ESC-50)}
The ESC-50 dataset~\cite{esc-50} is a widely used benchmark for environmental sound classification, comprising 2,000 labeled audio clips. It serves as an out-of-domain dataset to evaluate the generalization capability of the model. Following the experimental settings of EAT and M2D-CLAP, we employ a 5-fold cross-validation strategy and report the average accuracy (Acc) as the evaluation metric.

\noindent \textbf{Speech Commands (SPCV2)}
Speech Commands V2~\cite{spc-2} is a keyword spotting dataset consisting of 105,829 one-second audio recordings of spoken English words. The dataset is partitioned into training, validation, and test sets, containing 84,843, 9,981 and 11,005 samples, respectively. During fine-tuning, the model is trained on the training set, and classification accuracy (Acc) on the test set is used as the performance metric.

\section{Baseline Systems Detail}
To validate the effectiveness of our proposed method, we compare CAT with several representative state-of-the-art audio understanding models. These baselines cover three distinct training paradigms: supervised pretraining, contrastive learning pretraining, and self-supervised learning.

\subsection{Supervised Pretraining}
\textbf{AST}~\cite{ast} is a representative supervised learning model. It is the first attention-based model to introduce the Vision Transformer (ViT) architecture to the audio domain. AST processes audio spectrograms as a sequence of patch embeddings. During pre-training, it relies on large-scale supervised datasets (e.g., ImageNet for initialization followed by AudioSet) to learn audio representations. While effective, AST typically requires large amounts of labeled data, which limits its scalability compared to self-supervised approaches.

\subsection{Contrastive Learning Pretraining}
\textbf{CLAP} (Contrastive Language-Audio Pretraining)~\cite{clap} represents the contrastive learning paradigm. It consists of dual encoders—one for audio and one for text—trained jointly to align audio representations with their corresponding natural language descriptions. By utilizing large-scale text-audio pairs (e.g., LAION-Audio-630K), CLAP learns robust semantic representations capable of zero-shot transfer. In our work, we not only use CLAP as a strong baseline but also leverage its frozen audio encoder to provide representation regularization for our student model.

\subsection{Self-supervised Pretraining}
\textbf{AudioMAE}~\cite{audio-mae} is a generative self-supervised learning model. Following the Masked Autoencoder (MAE) framework from the vision domain , AudioMAE masks a high proportion of patches in the input spectrogram and trains the model to reconstruct the missing pixels of the spectrogram from the unmasked patches. This reconstruction-based objective encourages the model to learn local acoustic features and temporal dependencies without requiring labels.

\noindent \textbf{BEATs} (Bidirectional Encoder representation from Audio Transformers)~\cite{beats} is a state-of-the-art bootstrap-based self-supervised model. It employs an iterative pre-training strategy that utilizes an acoustic tokenizer to generate discrete labels for masked prediction. By learning to predict these discrete acoustic tokens, BEATs achieves superior performance on various audio understanding benchmarks. In our experiments, we compare against the $BEATs_{iter3}$ and $BEATs_{iter3+}$ variants.

\begin{table*}[htbp]
  \centering
  \caption{\textbf{Performance Comparison of Different Models on Audio Captioning Metrics.}}
  \label{tab:caption_performance}
  \begin{tabular}{lcccccc}
    \toprule
    Model & METEOR(\%) & CIDEr(\%) & SPICE(\%) & SPIDEr(\%) & SPIDEr-FL(\%) & FENSE(\%) \\
    \midrule
    \multicolumn{7}{l}{\textit{\textbf{Baseline model}}} \\
    BEATs~\cite{beats}                        & 18.5  & 48.8  & 13.3  & 31.0  & 29.6  & 50.1  \\
    Audio-MAE~\cite{audio-mae}                     & 15.6 & 31.2 & 10.5 & 20.8 & 20.8 & 41.8 \\
    AST~\cite{ast}                          & 8.9  & 8.1  & 4.0  & 6.0  & 6.0  & 21.4 \\
    
    \midrule
    \multicolumn{7}{l}{\textit{\textbf{Ours}}} \\
    CAT                & \textbf{19.8} & \textbf{50.7} & \textbf{14.9} & \textbf{32.8} & \textbf{32.6} & \textbf{53.1} \\
    \bottomrule
  \end{tabular}
\end{table*}

\section{Model Architecture Detail}
In the Multi-resolution Block, patch embedding is performed by a single 2D convolutional layer with a kernel size and stride both set to $r_{k+1}/r_k$. 
The number of channels gradually increases from 1 to the final dimension $D_{n_r}$, which is set to 768 to match the Transformer's hidden dimension.
The convolution module within the Multi-resolution Block consists of three sequential layers: a $1\times1$ convolution for cross-channel information aggregation, followed by a $5\times5$ convolution with a stride of 1 and padding of 2, and finally another $1\times1$ convolution. 
This design maintains both the feature map dimensions and the channel count.
The Downsample Block is implemented as a single convolutional layer with a kernel size and stride of $r_n/r_k$, which downsamples the feature map to match the dimensions of the final layer's features.

For the Transformer implementation, we use 12 layers for most ablation studies unless otherwise specified.
Since we additionally introduce the Multi-Resolution Block, we remove one layer in certain configurations to keep the number of trainable parameters roughly comparable across models, ensuring a fair comparison.
The attention module is configured with a hidden size of 768 and 12 heads.
Each Transformer block consists of a multi-head attention layer followed by an MLP block, which contains two fully-connected layers with an internal hidden dimension of $4 \times 768$.
The projector comprises five convolutional layers followed by a fully-connected layer. 
The first layer reduces the channel dimension from 768 to 384, while the subsequent four layers maintain this channel size. 
All convolutional layers use a $5 \times 5$ kernel with a padding of 2. 
Finally, a fully-connected layer projects the features back to the original dimension of 768.

\section{Training Hyper-Parameters}
Additional hyper-parameters used in pre-training using AS-2M and fine-tuning of standard audio SSL benchmark datasets are listed in Table~\ref{tab:hyperparams}.

\section{Model for Audio Caption}

To validate the performance of our model on other audio understanding tasks, we further evaluate the performance of our model and baseline models on the audio captioning task. 

\subsection{Audio Caption Steup}
We align the encoder representations to the language model’s (LM) representation space via a linear projector, which are then directly fed into the subsequent LM for audio captioning tasks.
We first pre-train the LM and then conduct supervised fine-tuning (SFT) — with LoRA~\cite{lora} applied throughout both stages to facilitate efficient parameter updates.
Following the experimental setup of SLAM-AAC~\cite{slamaac}, we perform pre-training on four datasets: Clotho~\cite{clotho}, AudioCaps~\cite{audiocaps}, WavCaps~\cite{wavcaps}, and MACS~\cite{macs}. 
We adopt paraphrasing augmentation following SLAM-AAC for data augmentation for Clotho dataset, which generates additional semantically consistent captions via back-translation to alleviate data scarcity. 
The pre-training data includes the training splits of Clotho, AudioCaps, and MACS, as well as the entire WavCaps dataset. 
Training is conducted with a global batch size of 16, a peak learning rate of 1e-4, a linear decay learning rate schedule with 1,000-iteration warmup, and a total of 100,000 training steps.
The model is further fine-tuned on the Clotho training set for 10 epochs (batch size=4, peak learning rate=8e-6), with its audio captioning performance evaluated on the Clotho evaluation split.
During inference, our model first generates multiple candidate captions following SLAM-AAC via beam search with varying beam sizes, then leverages the CLAP model to compute audio-text cosine similarity for each candidate and selects the one with the highest alignment score as the final output.

\subsection{Caption Results Analysis}

To assess the quality of generated audio captions, we evaluate the caption results using six standard Automated Audio Captioning (AAC) metrics: METEOR~\cite{meteor}, CIDEr~\cite{cider}, SPICE~\cite{spice}, SPIDEr~\cite{spider}, SPIDEr-FL~\cite{fence}, and FENSE~\cite{fence}.
METEOR evaluates unigram-level precision and recall, while incorporating synonyms and stemming to account for lexical variations. CIDEr quantifies the consensus between generated captions and references through TF-IDF weighted n-gram matching. SPICE takes a semantic graph-based approach, comparing the structural similarity of meaning representations between generated and reference texts. SPIDEr offers a balanced assessment by linearly combining CIDEr and SPICE, integrating both syntactic coherence and semantic relevance. Building on SPIDEr, SPIDEr-FL introduces fluency awareness by incorporating a fluency error detector from FENSE—a BERT-based model fine-tuned on audio captions—penalizing the SPIDEr score if the fluency error probability exceeds 90\%. FENSE itself provides a holistic evaluation by fusing Sentence-BERT-derived semantic similarity with fluency error detection.

As presented in Table~\ref{tab:caption_performance}, CAT consistently outperforms all baseline models across all reported metrics, demonstrating the superior semantic quality of its learned representations.
Specifically, CAT achieves a SPIDEr score of 32.8\%, surpassing the strong bootstrap-based baseline BEATs by 1.8\% and the generative Audio-MAE by a significant margin. Notably, the performance gap is even more pronounced on the SPIDEr-FL metric (+3.0\% over BEATs), indicating that the captions generated by CAT are not only semantically accurate but also more fluent.

We attribute this superior performance to the proposed Representation Regularization. CAT is explicitly aligned with CLAP, a model pre-trained on audio-text pairs. This alignment effectively injects rich, language-aligned semantic priors into the CAT encoder, enabling it to bridge the modality gap between audio signals and natural language descriptions more effectively than models trained solely on audio reconstruction or classification objectives.




\end{document}